\documentclass[aps,prl,superscriptaddress,prreprint]{revtex4-1}
 \usepackage{amsmath} \usepackage{amsfonts}
\usepackage{amssymb} 
\usepackage{array}
\usepackage{graphicx} 
\usepackage{verbatim}

\begin{document}

\section{Particle-Hole Symmetry  and \\  the Reentrant Integer Quantum Hall Wigner Solid}

Vidhi Shingla$^{1,\dag}$, Sean A. Myers$^{1,\dag}$, Loren N. Pfeiffer$^2$, Kirk W. Baldwin$^2$, and G\'abor A. Cs\'athy$^1$

$^1$ Department of Physics and Astronomy, Purdue University, West Lafayette, Indiana 47907, USA

$^2$ Department of Electrical Engineering, Princeton University, Princeton, New Jersey 08544, USA

$\dag$ These authors contributed equally

\subsection{Abstract}

The interplay of strong Coulomb interactions and of topology is currently under intense scrutiny in various
condensed matter and atomic systems. One example of this interplay
is the phase competition of fractional quantum Hall states and the Wigner solid
in the two-dimensional electron gas. Here we report a Wigner solid at  $\nu=1.79$
and its melting due to fractional correlations occurring at $\nu=9/5$. 
This Wigner solid, that we call the reentrant integer quantum Hall Wigner solid,
develops in a range of Landau level filling factors
that is related by particle-hole symmetry to the so called reentrant Wigner solid.
We thus find that the Wigner solid in the GaAs/AlGaAs system 
straddles the partial filling factor $1/5$ not only at the lowest filling factors, but also near $\nu=9/5$.
Our results highlight the particle-hole symmetry as a fundamental symmetry of the extended family of Wigner solids
and paint a complex picture of the competition of the Wigner solid with fractional quantum Hall states.

\subsection{Introduction}

One of the most stunning effects of strong interactions is the
formation of interaction-driven topological states, such as the fractional quantum Hall states (FQHSs) 
developing in clean two-dimensional electron gases (2DEGs) \cite{tsui}.
A fundamental symmetry that governs the fractional quantum Hall regime is the
particle-hole symmetry \cite{girvin,jain0,jain1}. While particle-hole symmetry is commonly applied
to the physics of the 2DEG, highly non-trivial aspects of this symmetry were not appreciated until recently.
Indeed, particle-hole symmetry near the Fermi sea of composite fermions was found
to have deep implications for the understanding their Fermi sea \cite{shay-phs}, explicit particle-hole symmetry was 
only recently demonstrated for transport and FQHS energy gaps \cite{pan-phs}, and the examination of particle-hole symmetry
for the even denominator FQHSs broadened concepts of topological order significantly \cite{ph1,ph2,ph3,ph4,ph5}.

Besides generating novel topological states, strong interactions in clean 2DEGs
also order charge carriers into electron solids \cite{ws-th1,ws-th2}. Due to the flat nature of energy bands
at high magnetic fields, one way for electrons to minimize their Coulomb energy is
through ordering into a triangular lattice called the Wigner solid (WS).
The insulating phase at Landau level filling factors $\nu < 1/5$ 
forming in the 2DEG in GaAs/AlGaAs heterostructures has long been interpreted as a WS,
often called the high field Wigner solid (HFWS) \cite{ws1,ws2,ws3,ws4,ws5}.
Results from a variety of experiments, such as non-linear transport \cite{ws6,ws7,ws8}, noise \cite{ws9},
dielectric constant measurement \cite{ws10}, microwave spectroscopy \cite{ws11,ws12,ws13},
composite fermion commensuration oscillations \cite{ws15}, and screening measurements \cite{ws16}
have strengthened this interpretation. Since the HFWS and the terminal FQHS at $\nu=1/5$ form in the 
same range of magnetic fields, these two phases often display a competition that determines
the structure of the phase diagram \cite{ws1,ws2,ws3,ws-th3,ws-th4}. 

Wigner crystallization in the 2DEG is energetically favored not only at $\nu < 1/5$,  
but also at other values of the filling factor. These Wigner solids bare different names.
For example, the Wigner solid that develops in the $1/5 < \nu < 2/9$ range of filling factors
is called the reentrant Wigner Solid (RWS) \cite{ws3}. Furthermore, the Wigner solid
referred to as integer quantum Hall Wigner solids (IQHWSs)
forms in the flanks of integer quantum Hall states (IQHS) \cite{int1,int2,int3,int4,int5,int6,int7,liu-1,liu-2}. 
IQHWSs have relatively large quasiparticle
densities at which the Anderson insulator in the bulk of the 2DEG reorganizes itself into
a Wigner solid once the Coulomb energy overtakes disorder effects.
IQHWSs were discovered by the detection of their pinning modes \cite{int1,int2,int3}, but were also observed 
in nuclear magnetic resonance \cite{int4}, compressibility \cite{int5}, 
surface acoustic wave propagation \cite{int6}, tunneling measurements \cite{int7},
and recent transport measurements \cite{liu-1,liu-2}. While the above WSs were discovered in GaA/AlGaAs hosts,
recent progress with the quality of other host materials yielded
similar physics. For example, Wigner crystallization has recently been observed in
the ZnO/ZnMgO system \cite{kozuka,maryenko}, graphene \cite{young}, and in AlAs quantum wells \cite{villegas}.

Here we report the observation of an electron solid 
near the filling factor $\nu=1.79$ in clean 2DEGs confined to GaAs/AlGaAs hosts.
The transport signatures of this phase, a vanishing magnetoresistance accompanied by
a Hall resistance that is quantized to $R_{xy}=h/2e^2$, allow us to
infer that this electron solid is a Wigner solid. 
The Wigner solid at $\nu=1.79$ develops near the IQHWS, but it is distinct from the latter. 
We therefore term this Wigner solid the reentrant integer quantum Hall Wigner solid (RIQHWS). 
We find that the RIQHWS is related by particle-hole symmetry to the RWS that develops at high magnetic fields.
This means that the elementary building blocks of the RIQHWS are quasiholes, rather than quasielectrons, 
and we establish that particle-hole symmetry is a fundamental symmetry that extends over the larger family 
of WSs. In addition, our observations reveal that the phase competition of the RIQHWS and a FQHS
that forms nearby is a generic property for 2DEGs with strong interactions which
has an imprint on the wider phase diagram.

\subsection{Results and Discussion}

Magnetotransport of our Sample 1 measured at $T=12$~mK reveals a rich structure exhibiting various known
ground states of the 2DEG. In Fig.1 we identify a few prominent examples,
such as the $\nu=2$, $3$, and $4$ IQHS, several FQHSs, including a rich structure
in the $N=1$ Landau level \cite{xia04,kumar10,ethan15}. In addition, the $\nu > 4$ range is dominated
by various electronic stripe and bubble phases \cite{fogler,moessner,lilly,du,eisen02}. 
We focus our attention on the range of filling factors
delimited by the $\nu=2$ IQHS and the $\nu=5/3$ FQHS, i.e. the region near $B=7.25$~T.
In this region in Fig.1 we observe structures in the longitudinal magnetoresistance $R_{xx}$. 

The region $5/3 < \nu < 2$ is shown in greater detail in Fig.2.
In the $T=300$~mK magnetoresistance $R_{xx}$ trace of Fig.2 there two local minima,
signaling developing FQHSs. One is at $\nu=nh/eB = 12/7$. Here $n$ is the electron density, $B$ is
the magnetic field, $e$ is the electron charge, and $h$ is Planck's constant. A resistance minimum at 
$\nu=12/7$ was attributed to a FQHS \cite{exp6}. In addition, in the $T=300$~mK $R_{xx}$ trace
there is another so far unreported resistance minimum at a higher filling factor $\nu=9/5$.
Hall resistance measurements at these two filling factors $\nu=12/7$ and $\nu=9/5$
reveal inflections close to $7h/12e^2$ and $5h/9e^2$, respectively.
These data are therefore suggestive of developing fractional correlations associated with 
FQHSs at $\nu=12/7$  and $\nu=9/5$. 
These and other Landau level filling factors of interest are marked in Fig.2 by dashed vertical lines.

\subsubsection{Observation of a collective insulator at $\nu=1.79$ and its competition with the $\nu=9/5$ FQHS}

Even though at $T=300$~mK we identified similar behavior at both $\nu=12/7$  and $\nu=9/5$, 
we find that the temperature evolution of the magnetoresistance at these two filling factors is
very different. Indeed, as the temperature is lowered from $T=300$~mK, 
the inflection in the Hall resistance at $\nu=12/7$
persists. As shown in Fig.2, at the lowest temperature of $T=12$~mK, $R_{xx}$ maintains its
local minimum and $R_{xy}$ is quantized to $7h/12e^2$. However, $R_{xx}$ at this filling factor
changes very little with temperature. This indicates a weak FQHS at $\nu=12/7$.
In contrast, the behavior near $\nu=9/5$ is qualitatively different.
The local minimum in $R_{xx}$ at $\nu=9/5$ is still present at $T=100$~mK, but it
rides on a magnetoresistance background that rises with lowering the temperature.
Indeed, as seen in Fig.2, $R_{xx}$ near $\nu=9/5$ at $T=100$~mK is approximately double of that at $T=300$~mK.
Furthermore, the local minimum in $R_{xx}$ at $\nu=9/5$ is extremely weak at $T=75$~mK
and it is no longer present at $T < 75$~mK.
We conclude that at temperatures $T \geq 75$~mK our sample exhibits
signs of fractional correlations at $\nu=9/5$. However,
at the lowest accessed temperatures our sample does not support a 
fractional quantum Hall ground state at $\nu=9/5$.

An unusual feature of the $T=75$~mK $R_{xx}$ data is that 
there is a local minimum in $R_{xx}$ which, however, 
is at a distinctively different filling factor than $9/5$. This new minimum in $R_{xx}$, 
marked by arrows in Fig.2, develops at $\nu=1.79$. This filling factor is not
a part of the standard sequence of fractional valued filling factors \cite{jain0},
therefore the minimum in $R_{xx}$ at $\nu=1.79$ most certainly cannot be associated with a FQHS.
This conclusion is strengthened by our observations of the
Hall resistance at $T \leq 75$~mK, which 
deviates significantly from any possible plateau values expected for a FQHS  
that may develop near $\nu=1.79$. Instead,
as the temperature is lowered, $R_{xx}$ at $\nu=1.79$ gradually decreases towards $R_{xx}=0$
and $R_{xy}$ becomes quantized to $R_{xy}=h/2e^2$. This is seen in the $T=23$~mK
and $T=12$~mK traces shown in Fig.2.
Such a behavior signals collective localization, i.e. the development of an electron solid \cite{lilly,du,eisen02}. 
In Fig.2 we marked  this electron solid at $\nu=1.79$ with arrows. Since a similar transport
signature is present along a perpedicular crystal direction, this electron solid is isotropic.

\subsubsection{Relating the electron solid at $\nu=1.79$ and the IQHWS}

In order to understand the nature of the electron solid forming at $\nu=1.79$, 
we recall that complex electron solids
called electronic bubble phases form in high Landau levels. Bubble phases are
a triangular lattice of electron bubbles, formed of electrons clustered together 
\cite{fogler,moessner,lilly,du,eisen02}. 
The clustering of several electrons into a bubble is afforded by the nodal structure
of the single electron wavefunction and it is consistent with the most recent
experimental observations \cite{zudov,kevin}. 
The electronic wavefunction in the $N=0$ Landau level does not have any nodes,
therefore among the electronic bubble phases only
the one-electron bubble phase may form,
a phase which is identical to the Wigner solid \cite{fogler,moessner}. 
Other charge ordered phases we have to consider are the ones constituted by
quasiparticles of FQHSs, such as the Wigner solid of composite fermions \cite{cfws1,cfws2,cfws3,cfws4}. 
However, because the Hall resistance is not quantized to a fractional value, in our experiment we have no
evidence of formation of such quasiparticles at $\nu=1.79$.
Since our electron solid forms in the $N=0$ Landau level, where multi-electron bubble
phases are not expected, and since at $\nu=1.79$ we have no
evidence of quasiparticles of FQHSs, we identify the solid at this filling factor with a Wigner solid.

As discussed in the introduction, Wigner solids termed IQHWSs are know to form in the flanks
of integer plateaus \cite{int1,int2,int3,int4,int5,int6,int7,liu-1,liu-2}, 
specifically in the flanks of the $\nu=2$ IQHS as well \cite{int1,int4}. 
IQHWSs form within the range of filling factors $0 <| \nu^* |< 1/5$,
albeit the stability range is expected to be temperature and disorder dependent 
\cite{int1,int2,int3,int4,int5,int6,int7,liu-1,liu-2}.
Here the partial filling factor near $\nu=2$ is $\nu^*=\nu-2$.
The partial filling factor of the $\nu=1.79$ Wigner solid is $|\nu^*|=0.21$,
which is clearly outside the range of formation of the IQHWS.
Hence the $\nu=1.79$ Wigner solid forms at a larger quasiparticle density than the nearby IQHWS. 
When a ground state becomes unstable for a small range of parameters, then reappears, it is called a reentrant
ground state. We will thus refer to the $\nu=1.79$ Wigner solid as the
reentrant integer quantum Hall Wigner solid (RIQHWS).

Before further analysis, we check that the filling factor of formation of the RIQHWS
is independent of experimental parameters we control.
We have already shown that the RIQHWS develops at a $T$-independent
$B$-field, and therefore a $T$-independent filling factor.
In addition, the RIQHWS also develops at a density-independent filling factor.
In Fig.3 we show the evolution of magnetoresistance with the density. Here we demonstrate that
as the density of Sample 1 is lowered from $3.05 \times 10^{11}$~cm$^{-2}$ to
$2.70 \times 10^{11}$~cm$^{-2}$, the RIQHWS develops at the same $\nu$.
Furthermore, the RIQHWS also develops at the same filling factor in
Sample 2 of a much reduced density $1.0 \times 10^{11}$~cm$^{-2}$. 
We think that the development of the RIQHWS in Sample 2 is aided by its
unusually high mobility and a wider width of the quantum well. 
The temperature and density independence of magnetoresistance
features associated with the RIQHWS establishes the RIQHWS as a genuine ground
state of the 2DEG over a wide range of electron densities.

\subsubsection{Relating the electron solid at $\nu=1.79$ and other WSs}

Further insight on the nature of the RIQHWS may be gained from 
considering particle-hole symmetry, a fundamental symmetry of the Hamiltonian of electrons in two dimensions
that connects ground states at different filling factors sharing the same absolute value of the partial filling 
factor $|\nu^*|$ \cite{girvin,jain0,jain1}. 
The $2-1/5 < \nu < 2$ range of stability of the IQHWS and
the range of stability $\nu < 1/5$ of the HFWS share the same 
range of partial filling factors $|\nu^*| < 1/5$. It is said that
the IQHWS and the HFWS are linked by the $\nu \leftrightarrow 2-\nu$ 
particle-hole symmetry, or HFWS~$\leftrightarrow$~IQHWS. 
We notice that the RIQHWS and the RWS are also linked by particle-hole
symmetry, i.e RWS~$\leftrightarrow$~RIQHWS. Indeed, the RIQHWS forms in the range
$9/5=2-1/5 < \nu < 2-2/9 =16/9$. The range of filling factors of $2-1/5 < \nu < 2-2/9$ the RIQHWS 
corresponds to partial fillings $1/5 < |\nu^*| < 2/9$ and it is indeed related by the same $\nu \leftrightarrow 2-\nu$  
symmetry to the $1/5<\nu<2/9$ range of the RWS \cite{ws3}. 
This means that the RWS is a WS of electrons, whereas the RIQHWS is a WS of hole quasiparticles
and both of these solids form in the same range of partial filling factors.
The stability ranges of the various WSs, including that of the RIQHWS, and their symmetries are on display in Fig.4.
All WSs marked in this figure straddle the partial filling factor $|\nu^*|=1/5$ and obey particle-hole symmetry.
As a consequence, we found that the RIQHWS, the newest member of the WS family, together
with other WSs, admit particle-hole symmetry as a fundamental property.

Our results find a natural explanation if one assumes that the IQHWS and the RIQHWS 
are part of the same monolithic WS phase. This monolithic WS phase straddles the filling factor $\nu=9/5$,
i.e. it is present on both sides of $\nu=9/5$.
The resistive peak in $R_{xx}$ seen in at $\nu=9/5$ Fig.3 and also in Fig.2 at the four lowest temperatures 
is then due to melting of this WS due to remnant fractional correlations at this filling factor. 
The observed magnetotrasport thus indicates a competition of two strongly correlated ground states: the
WS and the $\nu=9/5$ FQHS. Invoking 
such remnant correlations at $\nu=9/5$ is not unreasonable since their effect
may be observed in Fig.2 as an incipient FQHS at $75 \leq T \leq 300$~mK. These fractional correlations
in our sample are, however, not strong enough to allow the development of a fully quantized 
fractional quantum Hall ground state at $\nu=9/5$. Indeed, according to data plotted in Fig.3,
$R_{xx}(\nu=9/5) \neq 0$ and $R_{xy}(\nu=9/5) \neq 5h/9e^2$ for both samples.
$R_{xy}$ at $\nu=9/5$ has a sholder in Sample 2, but $R_{xy}(\nu=9/5)$
is $3.4\%$ less than $5h/9e^2$. A similar situation was encountered 
in early experiments on low mobility 2DEGs in which insulating behavior was seen on both sides
of $\nu=1/5$, i.e. in which both the HFWS and the RWS were present, but a FQHS at $\nu=1/5$
between these WSs did not fully develop \cite{ws1,ws2}. Therefore our results show that
the competition of WSs and the FQHS that they straddle is a generic feature of phase competition
in regions where WSs form.

The WSs discussed so far form in the $N=0$ orbital Landau level, i.e. in the $0 < \nu <2$ range.
We suggest that the above properties of the WS may be further extended even into the $N=1$
Landau level. The IQHWS was known within the $2 < \nu < 2+1/5$ range \cite{int1,int4}
and in an earlier publication a WS was found 
at $\nu=2.21$, clearly confined to the $2+1/5 < \nu < 2+2/9$ range \cite{deng-2,gabor}.
For this WS $1/5 < |\nu^*| < 2/9$ and,
based on our arguments put forth earlier, we identify this WS at $\nu=2.21$ with the RIQHWS.
Furthermore, there is a monolithic WS phase in this region that straddles both sides of $\nu=2+1/5$
FQHS and which is melted by the fractional correlations of this FQHS. This RIQHWS
is related by particle-hole symmetry to both the RIQHWS at $\nu=1.79$ and to the RWS,
and it is included in the diagram shown in Fig.4.
We note that in Sample 1 this WS at $\nu=2.21$ did not develop. The reason for this is unknown,
but differences in density, mobility, different measurement temperatures, 
and a different background impurity profile due to the diffferent growth chambers are possible culprits.

Most recently electron solids in the $N=0$ Landau level were also reported close to $\nu=1$ \cite{liu-1,liu-2}.
Some of these are in the range of filling factors associated with the IQHWS.
Others, such as the ones forming at
$\nu=0.78$ in a $42$~nm, $44$~nm \cite{liu-1}, and  a $65$~nm quantum sample \cite{liu-2}
develop at partial filling factor $|\nu^*|=0.22 $.
These electron solids were first interpreted as WSs \cite{liu-1}, then later as exotic solids formed of composite
fermions \cite{liu-2}. The partial filling factor of these electron solids is stunningly close to
that of the RIQHWS we found. One possibility is thus that these WSs are identical to the RIQHWS, particle-hole
symmetric counterparts of the RWS. Nonetheless, the nature of these electron solids remains to be determined;
softening of the short range part of the effective electron-electron interaction that occurs in samples with
wide quatum wells, for example, is known to tune the formation of electron solids and may stabilize various
types of electronic solids.

A shared feature of the WSs in our electron samples is that they straddle the partial filling factor $|\nu^*|=1/5$.
In other systems, however, WSs may straddle other partial filling factors. 
For example, in two-dimensional hole gases (2DHGs) the WS straddles $\nu^*=\nu=1/3$ \cite{santos,ccli,rui}.
This difference between 2DEGs and 2DHGs is attributed to the significantly larger ratio of the Coulomb interaction
and Fermi energies in the latter \cite{santos,ccli}.
Furthermore, the WS in 2DEGs in very narrow quantum wells straddles $\nu=1/3$ \cite{woowon} 
and WSs form in samples with added short-range disorder near $|\nu^*|=1/3$ \cite{luhman}.
In yet other systems, phases associated with the WS may straddle more than just one FQHS.
This is the case of a 2DEG realized in the ZnO/MgZnO system in which the WS 
was found to straddle several FQHSs \cite{maryenko}.
These experiments suggest that disorder influences the range of stability of the WS and
highlight that there is still much to be understood about disorder effects.

\subsection{Conclusions}

We reported complex features of the magnetoresistance at $\nu=1.79$ in the $N=0$ Landau level of two
high mobility 2DEGs confined to GaAs/AlGaAs. In contrast to similar transport features 
in high $N \geq 2$ Landau levels, the ground state at $\nu=1.79$ is associated with a Wigner solid.
Because of its proximity to the IQHWS, we named this WS the RIQHWS.
Based on the stability region in the filling factor space,  the RIQHWS and another RIQHWS reported 
at $\nu=2.21$ in the $N=1$ Landau level \cite{deng-2,gabor},
can both be understood as the particle-hole symmetric counterparts of the RWS. 
Our results indicate that WSs in the GaAs/AlGaAs system straddle the partial filling factor $|\nu^*|=1/5$,
that particle-hole symmetry of the WS is more pervasive than previously
thought, and that this symmetry leads to competing Wigner solids and fractional quantum Hall states
in unexpected parts of the phase diagram.

\section{Methods}

\subsection{Samples}
Sample 1 is a 2DEG confined to a 30~nm GaAs quantum well that is part of a GaAs/AlGaAs heterostructure.
The density of this sample is $n=3.05 \times 10^{11}$~cm$^{-2}$ and low temperature mobility is
$\mu=3.2 \times 10^7$~cm$^2$/Vs. The 2DEG is doped in a superlattice. The sample state was prepared 
with illumination with a red light emitting diode at
$10$~K according to a procedure described in Ref.\cite{light2}. 

Sample 2 is also a 2DEG confined to a GaAs/AlGaAs heterostructure, but differs in several aspects from Sample 1. 
It belongs to the most recent generation of high mobility samples \cite{loren}. It density is significantly
less than that of Sample 1, $n=1.0 \times 10^{11}$~cm$^{-2}$, its mobility is
$\mu=3.5 \times 10^7$~cm$^2$/Vs. The width of the confining quantum well in Sample 2 is 49~nm,
significantly larger than that of Sample 1. We note that 
since the mobility of Sample 2 is very large, the second electric subband in this sample is not populated.

\subsection{Experimental techniques}
Magnetotransport measurements were performed in a van der Pauw sample geometry using
standard lock-in technique. The square shaped samples have 8 indium ohmic contacts placed 
in the corners and the middle of the sides. The excitation current used was $3$~nA.
Sample 1 was mounted in vacuum on the copper tail of our dilution refrigerator,
reaching the lowest estimated temperature of $T = 12$~mK. 
In contrast, Sample 2 was mounted inside a He-3 immersion cell,
with the lowest bath temperature reached in our current measurement being $T=5.1$~mK \cite{setup}.
In this setup the electronic temperature follow closely that of the bath \cite{setup}.
For temperature measurements a carbon thermometer was used \cite{thermo} which below $100$~mK
was calibrated against a He-3 quartz tuning fork viscometer \cite{setup}. 

The density of Sample 1 was reduced from its maximum value 
by using a cold illumination technique described in Ref.\cite{light2}. 
As a result of this process, sample parameters changed from
of $3.05 \times 10^{11}$~cm$^{-2}$ and $\mu=3.2 \times 10^7$~cm$^2$/Vs to
to $2.70 \times 10^{11}$~cm$^{-2}$ and $\mu=1.3 \times 10^7$~cm$^2$/Vs.
We passed bursts of current of  $1$~$\mu$A through a red light emitting diode
for a duration of about $5$~s, while the sample was kept near the base temperature
of the refrigerator and at $B=0$. For the illumination procedure to work it is critical that the light beam 
covers the area of the sample uniformly. To achieve this, the diode faces the sample, it is centered
onto the sample, and it is mounted at a distance of $1.5$~cm from the sample.

\section{Acknowledgements}

Measurements at Purdue were supported by the NSF DMR Grant No. 1904497. The sample growth effort of 
L.N.P. and K.W.B. of Princeton University was supported by the Gordon and Betty Moore Foundation 
Grant No. GBMF 4420, and the National Science Foundation MRSEC Grant No. DMR-1420541.

{\bf Author contributions.}
V.S. and S.A.M. performed low temperature transport measurements. L.N.P. and K.W.B. produced MBE grown
GaA/AlGaAs samples and characterized them. V.S., S.A.M., and G.A.C. conceived the project, analyzed
the data, and wrote the manuscript.
\newline

\begin{figure}
\includegraphics[width=1\columnwidth]{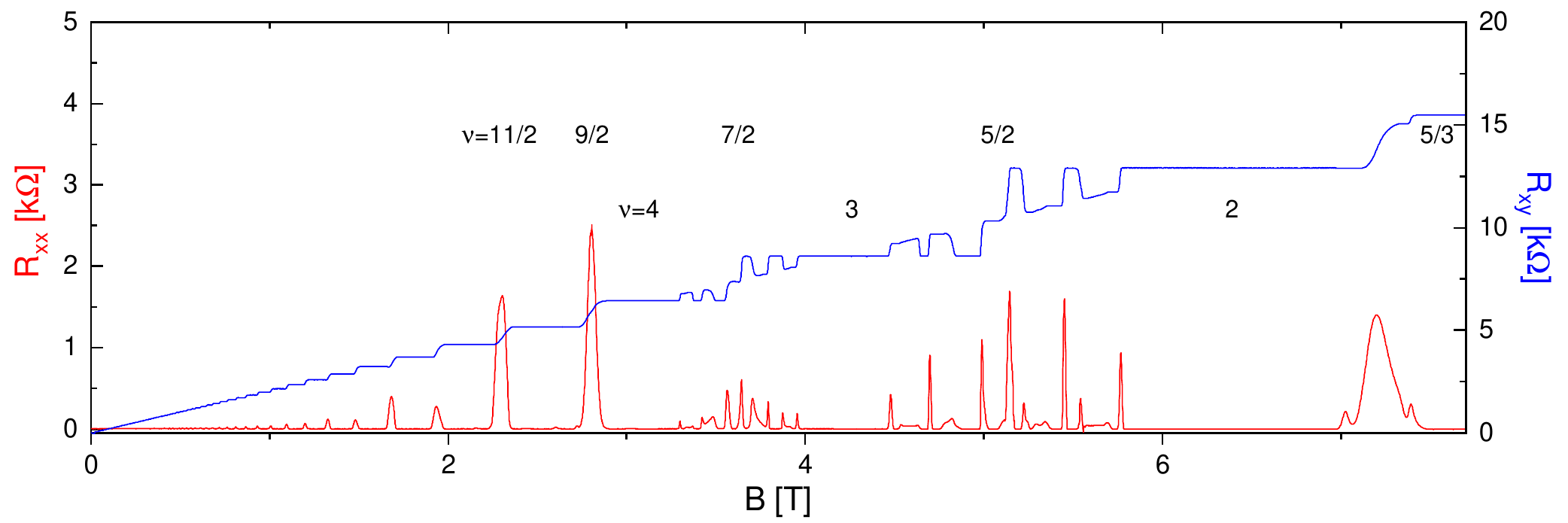}
\caption{
\textbf{Magnetoresistance $R_{xx}$ and Hall resistance $R_{xy}$ of Sample 1 over a broad range of
magnetic fields $B$.} Traces are obtained at the temperature $T=12$~mK. 
Numerical labels indicate notable Landau level filling factors $\nu$. 
Structures of interest develop near $B=7.25$~T, in the range of filling factors $5/3 < \nu < 2$.
} 
\end{figure}

\begin{figure}
\includegraphics[width=0.9\columnwidth]{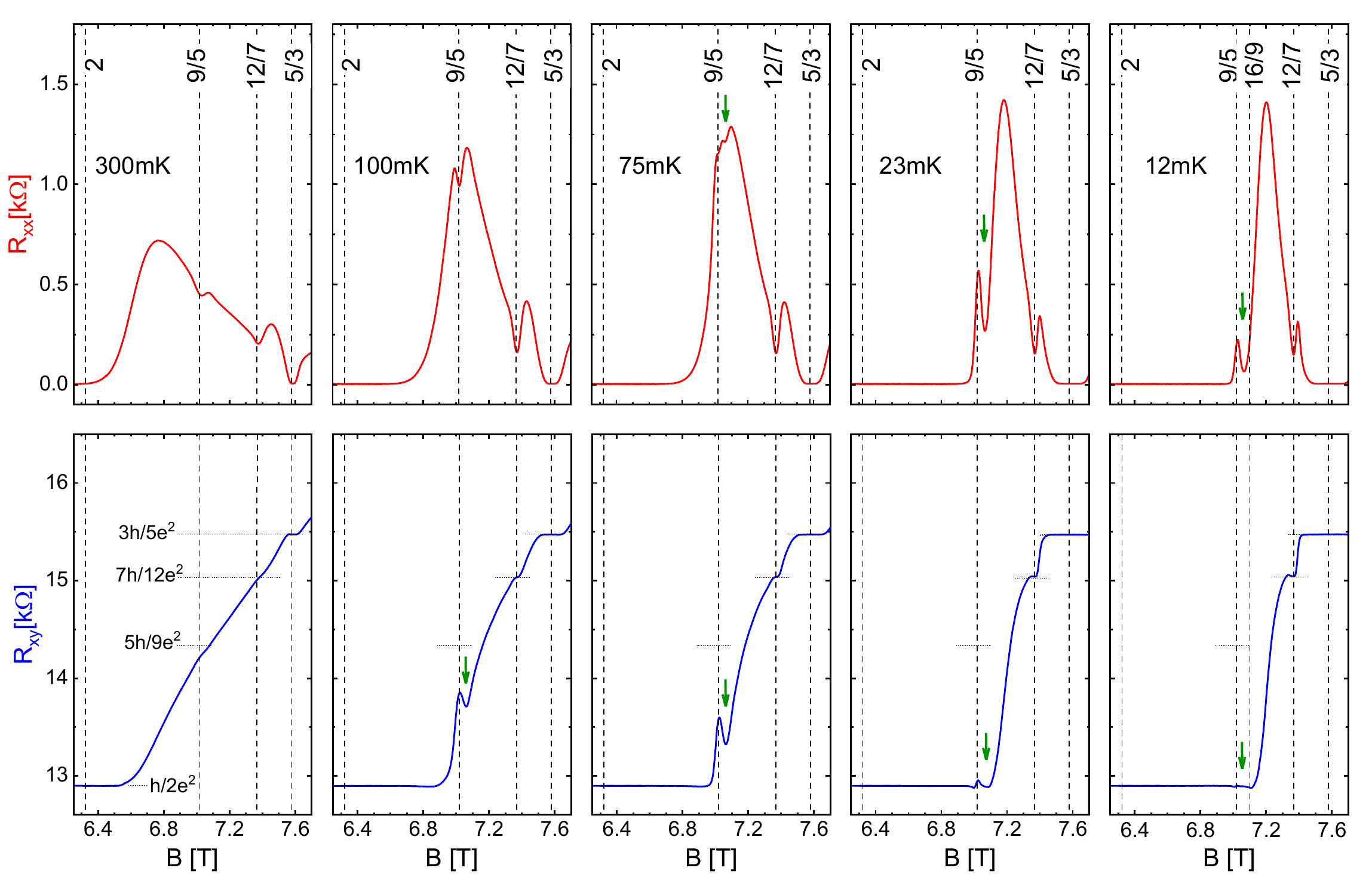}
\caption{
\textbf{The dependence on temperature $T$ of the magnetoresistance $R_{xx}$ 
and Hall resistance $R_{xy}$ of Sample 1
in the range of filling factors $5/3 < \nu <  2$.} Labels on vertical dashed lines are filling factors $\nu$ of interest.
The electron solid we refer to as the reentrant integer quantum Hall Wigner solid (RIQHWS) is
located between filling factors $\nu=9/5$ and $16/9$ and it is marked by arrows. 
} 
\end{figure}

\begin{figure}
\includegraphics[width=0.8\columnwidth]{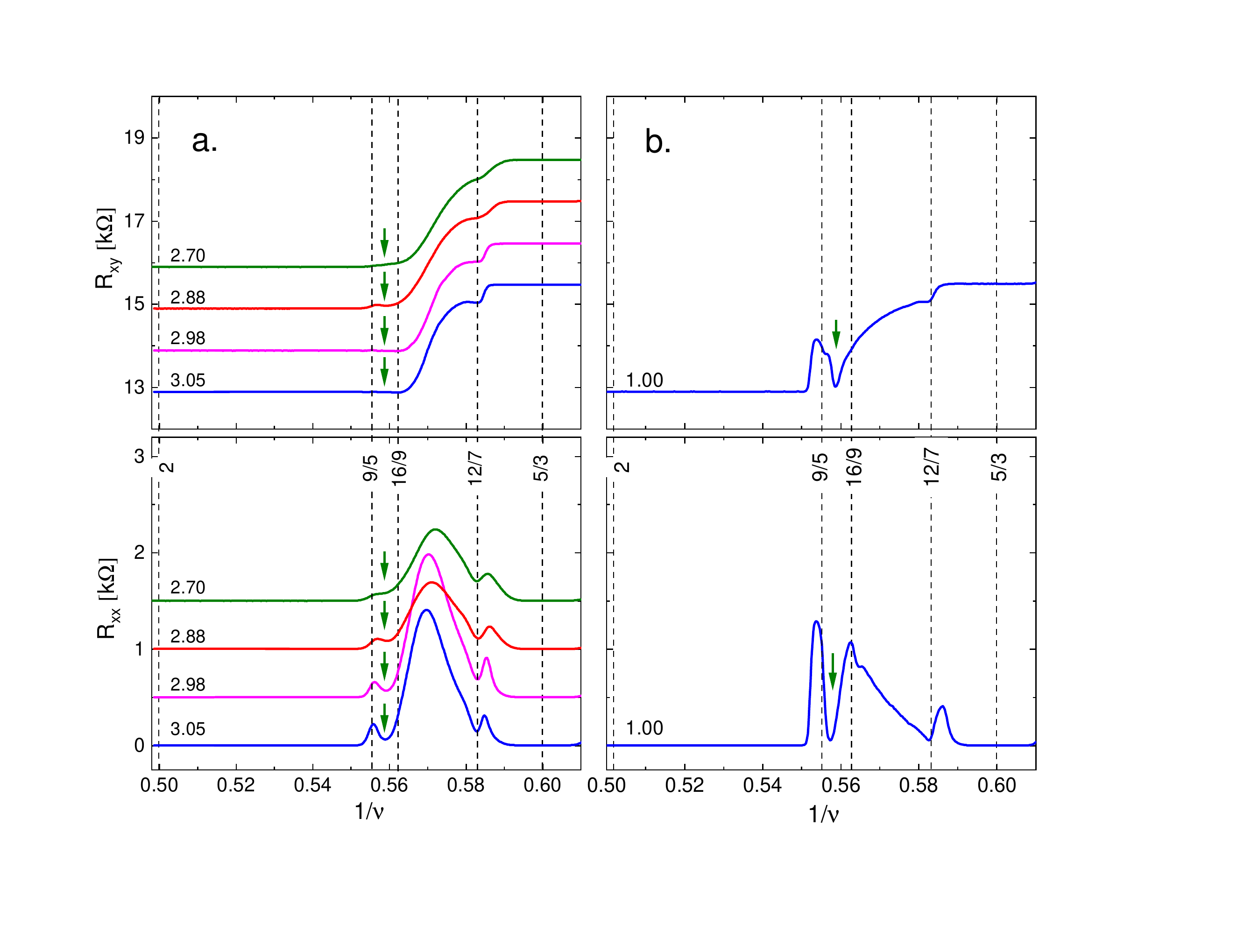}
\caption{
\textbf{Reentrant integer quantum Hall Wigner solids in a variable density sample and a sample of fixed density.}
Panel a.: Waterfall plots of the magnetoresistance $R_{xx}$ and 
Hall resistance $R_{xy}$ as plotted against the inverse filling factor $1/\nu$ of Sample 1
as prepared at different electron densities. Measurements were performed at the temperature of $T = 12$~mK. 
Panel b: Magnetoresistance $R_{xx}$ and Hall resistance $R_{xy}$ 
as plotted against the inverse filling factor $1/\nu$ of Sample 2
of a fixed electron density. This density is significantly lower than that of Sample 1.
Measurements were performed at the temperature of  $T = 5.1$~mK. Labels on traces indicate electron densities $n$
in units of $10^{11}$~cm$^{-2}$. Vertical dashed lines mark locations and values of
filling factors $\nu$ of interest and arrows mark the reentrant integer quantum Hall Wigner solid (RIQHWS). 
With the exception of traces at the highest density, traces for Sample 1 are shifted vertically upward for clarity.} 
\end{figure}

\begin{figure}
\includegraphics[width=0.8\columnwidth]{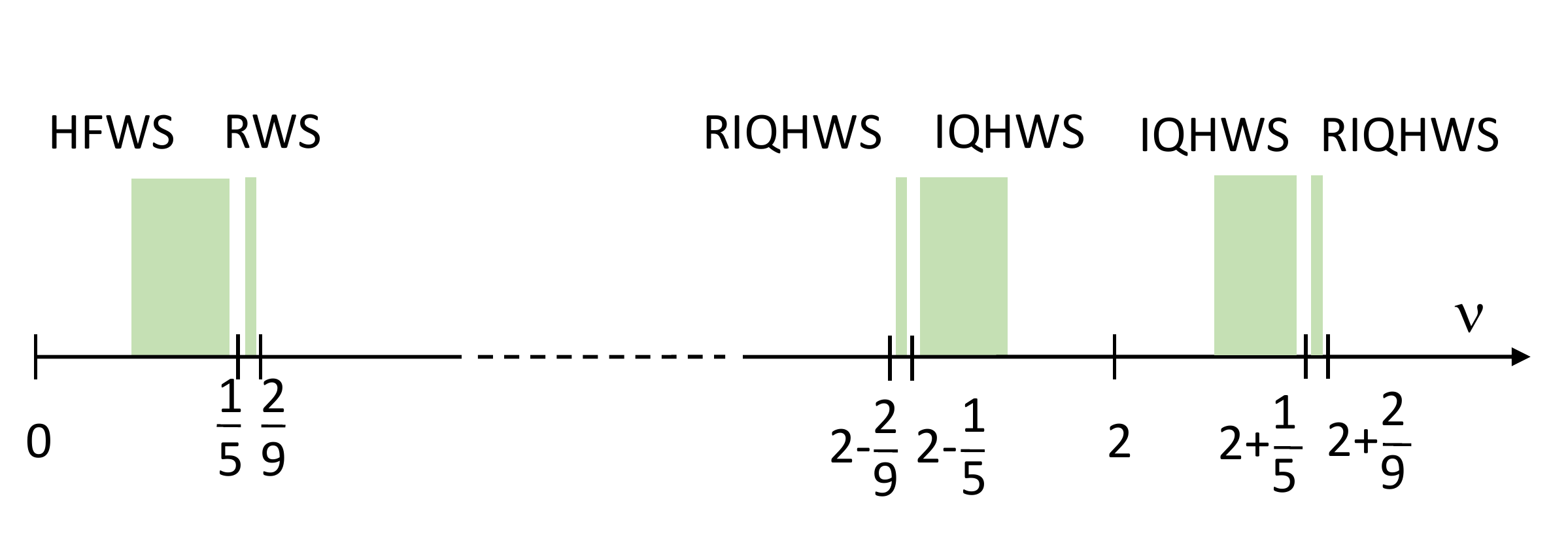}
\caption{
\textbf{Schematic diagram of the stability regions of the various Wigner solids in GaAs/AlGaAs system.}
  The Wigner solid we report on is the reentrant integer quantum Hall Wigner solid (RIQHWS), seen 
  between the filling factors $\nu=2-1/5$ and $\nu=2-2/9$.
  The high field Wigner solid (HFWS) and integer quantum Hall Wigner solid (IQHWS) 
  are related by particle-hole symmetry. Similarly, the reentrant Wigner solid (RWS) and the 
  reentrant integer quantum Hall Wigner solid (RIQHWS) are also related by particle-hole symmetry.} 
\end{figure}

\end{document}